\documentclass[showpacs,aps,graphicx,twocolumn]{revtex4}

\usepackage{graphicx}

\begin{document}


\title{Hyperentanglement purification and concentration assisted by diamond NV centers inside
photonic crystal cavities\footnote{Published in Laser Physics
Letters \textbf{10}, 115201 (2013)} }

\author{Bao-Cang Ren and Fu-Guo Deng\footnote{Corresponding author: fgdeng@bnu.edu.cn} }

\address{ Department of Physics, Applied Optics Beijing Area Major Laboratory,
Beijing Normal University, Beijing 100875, China}

\date{\today }

\begin{abstract}
Hyperentanglement has attracted much attention due to its
fascinating applications in quantum communication. However, it is
impossible to purify a pair of photon systems in a mixed
hyperentangled state with errors in two degrees of freedom using
linear optical elements only, far different from all the existing
entanglement purification protocols in a degree of freedom (DOF) for
quantum systems. Here, we investigate the possibility of purifying a
spatialpolarization mixed hyperentangled Bell state with the errors
in both the spatial-mode and polarization DOFs, resorting to the
nonlinear optics of a nitrogen-vacancy (NV) center in a diamond
embedded in a photonic crystal cavity coupled to a waveguide. We
present the first hyperentanglement purification protocol for
purifying a pair of two-photon systems in a mixed hyperentangled
Bell state with the errors in two DOFs. We also propose an efficient
hyperentanglement concentration protocol for a partially
hyperentangled Bell pure state, which has the maximal success
probability in principle. These two protocols are useful in
long-distance quantum communication with hyperentanglement.
\end{abstract}

 \maketitle

\section{\uppercase\expandafter{1.} Introduction}

Entanglement, a unique phenomenon in quantum physics, is an
essential resource in quantum information processing, such as
quantum computing \cite{QC}, quantum key distribution \cite{QKD},
quantum secure direct communication \cite{QSDC1,QSDC2},  quantum
teleportation \cite{QT1}, quantum dense coding \cite{DC1,DC2},
quantum secret sharing \cite{QSS1,QSS2}, and so on. In long-distance
quantum communication,  entangled photon pairs always act as a
quantum channel and they are usually produced locally and inevitably
polluted by the environment noise in the distribution process
between the two legitimate users. With current technology, the
photon signals can be transmitted no more than several hundreds of
kilometers in an optical fiber or a free space, so  quantum
repeaters are required to connect the two neighboring nodes in a
long-distance quantum communication network.  Also, the entanglement
of photon pairs will decrease in the storage process, which will
decrease the  fidelity and the security of long-distance quantum
communication protocols. In quantum repeaters, entanglement
purification and entanglement concentration are two necessary
quantum techniques for improving the entanglement of the entangled
systems.

Entanglement purification is used to obtain high-fidelity nonlocal
entangled states from mixed states with less entanglement
\cite{EPP1,EPP2,EPP3,EPPsimon,EPPexperiment,EPPsheng1,EPPsheng2,
EPPsheng3,EPPLi,EPPdeng1}. Up to now, many entanglement purification
protocols (EPPs) were proposed for the polarization photon systems,
resorting to nonlinear optics \cite{EPP1,EPP2,EPPsheng1,EPPsheng2}
or linear optics
\cite{EPP3,EPPsimon,EPPexperiment,EPPsheng3,EPPLi,EPPdeng1}.
Entanglement concentration is used to obtain the maximally entangled
states from partially entangled pure states
\cite{ECP1,ECP2,ECP7,ECP8,ECP4,ECP5,ECP6,HECP}. By far, some
important entanglement concentration protocols (ECPs) were proposed
for the polarization entanglement of photon systems with the
parameters of the partially entangled pure state either accurately
known \cite{ECP1,ECP2,ECP7,ECP8,HECP} or unknown
\cite{HECP,ECP4,ECP5,ECP6} to the two remote users in quantum
communication.

Hyperentanglement, defined as the entanglement of a quantum system
in multiple degrees of freedom
\cite{GenH,hyperentanglement,hypercnot}, has attracted much
attention for its fascinating applications in quantum information
processing. It can be used to assist polarization entanglement
purification
\cite{EPPsheng2,EPP3,EPPsimon,EPPexperiment,EPPsheng3,EPPLi,EPPdeng1},
polarization  Bell-state analysis
\cite{BSA1kwiat,BSA2walborn,BSA3,BSA4,BSA} and polarization quantum
repeater \cite{wangrepeater}.  In 2008, Barreiro \emph{et al}.
\cite{HESC} beat the channel capacity limit of superdense coding
with polarization-orbital-angular-momentum hyperentanglement using
linear optics. Recently, some hyperentangled Bell-state analysis
protocols were proposed with nonlinear optics for increasing the
channel capacity of long-distance quantum communication
\cite{kerr,HBSA,HBSA1}.

Although there are some interesting entanglement purification  and
entanglement concentration protocols, they are focused on only a
degree of freedom  (DOF) of quantum systems. There is no
hyperentanglement purification protocol (hyper-EPP). Moreover, it is
impossible to complete the hyperentanglement purification for a pair
of quantum systems entangled in two or more DOFs with linear optical
elements, far different from the existing EPPs
\cite{EPP3,EPPsimon,EPPexperiment,EPPsheng3,EPPLi,EPPdeng1}. Here,
we investigate the possibility of achieving the hyperentanglement
purification of two-photon systems in nonlocal hyperentangled Bell
states in both the spatial mode and   polarization DOFs with the
parity-check quantum nondemolition detectors (QNDs) that are
constructed by the nonlinear optics of a nitrogen-vacancy (NV)
center in a diamond embedded in a photonic crystal cavity coupled to
a waveguide. We propose the first hyper-EPP for a
spatial-polarization mixed hyperentangled Bell state with the errors
in both the polarization and  spatial mode DOFs.  Also, we propose
an efficient hyperentanglement concentration protocol (hyper-ECP)
which can be used to concentrate the partially hyperentangled Bell
states with unknown parameters. With the iteration of our hyper-ECP
process, the maximal success probability can be achieved in
principle, which is much higher than that with linear optics.

\section*{\uppercase\expandafter{2.} Hyper-EPP for a mixed hyperentangled Bell state with cavity-NV-center systems}
\label{sec2}

\subsection*{\uppercase\expandafter{2.1} Parity-check QNDs for the
polarization and  spatial-mode DOFs of two-photon systems}
\label{sec21}

\begin{figure}[htbp]             
\centering\includegraphics[width=7.6 cm]{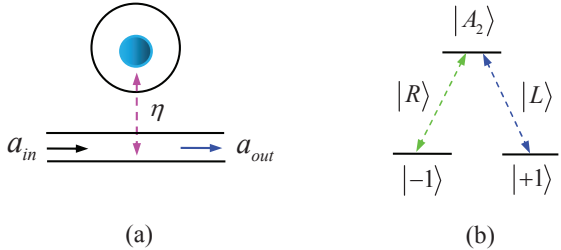} \caption{The
optical transition  of an NV-center  embedded in a photonic crystal
cavity with circularly polarized photons. (a) One-side
cavity-NV-center (cavity-waveguide-NV-center) system. (b) The  spin
preserving  optical transition  between the ground state and the
excited state of an NV center.} \label{figure_1}
\end{figure}

The schematic diagram for an NV center in a diamond embedded in a
photonic crystal cavity coupled to a waveguide is shown in
Fig.\ref{figure_1}. The negatively charged  NV  center is consisted
of a substitutional nitrogen atom and an adjacent vacancy with six
electrons from the nitrogen and three carbons surrounding the
vacancy. The ground state is an electronic spin triplet with a
splitting of $2.88$ GHZ between the magnetic sublevels $|0\rangle$
($m_s=0$) and $|\pm1\rangle$ ($m_s=\pm1$), and its orbit state is
$|E_0\rangle$. The specifically excited state
$|A_2\rangle=\frac{1}{\sqrt{2}}(|E_-\rangle|+1\rangle+|E_+\rangle|-1\rangle)$
\cite{NV2}, which is induced by spin-orbit and spin-spin
interactions and $C_{3v}$ symmetry \cite{NV4}, is robust with the
stable symmetry. It decays with an equal probability to the two
ground states $|-1\rangle$ and $|+1\rangle$ with radiation of right
($|R\rangle$, $\sigma_+=+1$) and left ($|L\rangle$, $\sigma_-=-1$)
circularly polarized photons, respectively. Here, $|E_\pm\rangle$
($|E_0\rangle$) are the orbit states with the angular momentum
projections $\pm1$ ($0$) along the NV axis.

The input-output property of a cavity-NV-center system shown in Fig.
\ref{figure_1} (a) can be described by the Heisenberg equations of
motion for the cavity field operator $\hat{a}$ and diploe operator
$\hat{\sigma}_-$ \cite{QD11},
\begin{eqnarray}                           
\frac{d\hat{a}}{dt}&=&-\left[i(\omega_c-\omega)+\frac{\eta}{2}+\frac{\kappa}{2}\right]\hat{a}-
g\,\hat{\sigma}_{-}
- \sqrt{\eta}\,\hat{a}_{in}, \nonumber\\
\frac{d\hat{\sigma}_-}{dt}&=&-\left[i(\omega_e-\omega) +
\frac{\gamma}{2}\right]\hat{\sigma}_{-} - g\,\hat{\sigma}_z\hat{a}.
\end{eqnarray}
where $\omega_e$, $\omega$,  and $\omega_c$ are the frequencies of
the energy transition, the waveguide channel mode,  and the cavity
mode, respectively. $g$ is the coupling strength between the cavity
and the NV center. $\gamma/2$, $\eta/2$,  and $\kappa/2$ are the
decay rates of the NV center, the cavity field,  and the cavity side
leakage mode, respectively. $\hat{a}_{in}$ and $\hat{a}_{out}$ are
the input and output field operators of the waveguide, and they
satisfy the relation
$\hat{a}_{out}=\hat{a}_{in}+\sqrt{\eta}\,\hat{a}$. The reflection
coefficient ($\hat{a}_{out}/\hat{a}_{in}$) of the cavity-NV-center
system can be obtained in the weak excitation limit (the NV center
is dominantly in the ground state and $\langle\sigma_z\rangle=-1$)
\cite{CNV1,CNV2} , that is,
\begin{eqnarray}                           
r(\omega)&=&\frac{\left[i(\omega_e-\omega)+\frac{\gamma}{2}\right]\left[i(\omega_c-\omega)
-\frac{\eta}{2}+\frac{\kappa}{2}\right]+g^2}{\left[i(\omega_e-\omega)
+\frac{\gamma}{2}\right]\left[i(\omega_c-\omega)+\frac{\eta}{2}+\frac{\kappa}{2}\right]+g^2}.\;\;\;\;
\end{eqnarray}

\begin{figure}[!h] 
\centering
\includegraphics[width=8.0 cm]{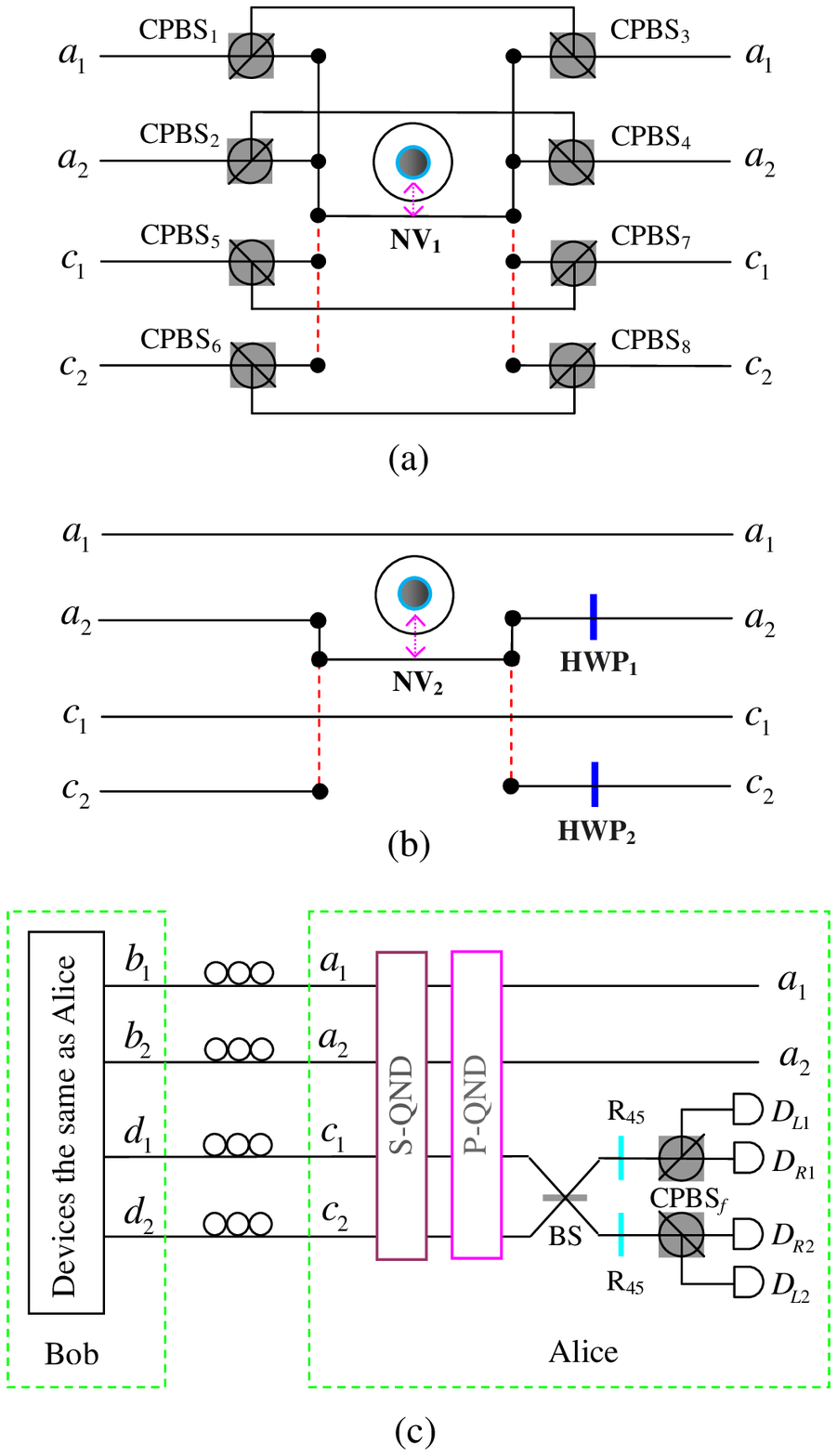}
\caption{(a) Schematic diagram of the polarization parity-check QND
(P-QND). (b) Schematic diagram of the spatial-mode parity-check QND
(S-QND).  (c)  Schematic diagram of our hyper-EPP for a mixed
hyperentangled Bell state with both spatial-mode bit-flip errors and
polarization bit-flip errors,  resorting to S-QNDs and P-QNDs. Bob
performs the same operations as Alice by replacing the photons $A$
and $C$ with the photons $B$ and $D$, respectively. NV$_1$ (NV$_2$)
represents a one-side cavity-NV-center system which is used to
perform polarization (spatial-mode) parity-check QND on the photon
pair $AC$. CPBS$_i$ ($i=1,2,\cdots$) represents a polarizing beam
splitter in the circular basis, which transmits the photon in the
right-circular polarization $|R\rangle$ and reflects the photon in
the left-circular polarization $|L\rangle$, respectively. $R_{45}$
represents a half wave plate which is used to perform a Hadamard
operation on the polarization DOF of the photon. HWP$_1$ (HWP$_2$)
represents a half wave plate which is used to perform a polarization
phase-flip operation $\sigma^p_z=|R\rangle\langle
R|-|L\rangle\langle L|$. $D_j$ ($j=L1,R1,R2,L2$) represents a
single-photon detector. BS represents a 50:50 beam splitter which is
used to perform a Hadamard operation on the spatial-mode DOF of the
photon. $k_1$ and $k_2$ represent the two spatial modes of the
photon $k$ ($k=a,c,b, d$).} \label{figure2}
\end{figure}

In an ideal condition, the cavity side leakage can be neglected, and
the NV center is resonant with the cavity
($\omega_c=\omega_e=\omega_0$). If the NV center is uncoupled to the
cavity ($g=0$), the reflection coefficient is $r_0(\omega)=-1$. If
the NV center is coupled to the cavity ($4g^2\gg\eta\gamma$), the
reflection coefficient is $r(\omega)\rightarrow1$. That is,
\begin{eqnarray}                           
&&|R,-1\rangle \;\rightarrow\;\;\; |R,-1\rangle,\;\;\;\;
|R,+1\rangle \;\rightarrow\;
-|R,+1\rangle,\nonumber\\
&&|L,-1\rangle \;\rightarrow\; -|L,-1\rangle,\;\;\;\; |L,+1\rangle
\;\rightarrow\;\;\;\; |L,+1\rangle.
\end{eqnarray}
This input-output property of a cavity-NV-center system can be used
to construct parity-check QNDs for both the polarization (NV$_1$ in
Fig. \ref{figure2} (a)) and the spatial-mode (NV$_2$ in Fig.
\ref{figure2} (b)) DOFs of a two-photon system. Suppose that the
initial states of NV$_1$ and NV$_2$ are
$\frac{1}{\sqrt{2}}(|-1\rangle+|+1\rangle)_{e_1}$ and
$\frac{1}{\sqrt{2}}(|-1\rangle+|+1\rangle)_{e_2}$, respectively.

(1) \emph{Polarization parity-check QND.} The polarization
parity-check QND (P-QND) is constructed with CPBSs and NV$_1$ shown
in Fig.\ref{figure2} (a). If the polarization DOF of the two-photon
system $AC$ is in a Bell state, we first let the photon $A$ pass
through CPBS (CPBS$_1$ and CPBS$_2$), NV$_1$, and CPBS (CPBS$_3$ and
CPBS$_4$), and then put the photon $C$ into CPBS (CPBS$_5$ and
CPBS$_6$), NV$_1$, and CPBS (CPBS$_7$ and CPBS$_8$).  After
interaction, the state of the system composed of  NV$_1$ and  two
photons with  polarizations  becomes
\begin{eqnarray}                           
|\phi^\pm\rangle^P_{AC} \otimes |+\rangle_{e_1}
 &\rightarrow& |\phi^\pm\rangle^P_{AC}
\otimes  |+\rangle_{e_1},\nonumber\\
|\psi^\pm\rangle^P_{AC} \otimes |+\rangle_{e_1}
 &\rightarrow& |\psi^\pm\rangle^P_{AC} \otimes
 |-\rangle_{e_1}.\;\;\;\;
\end{eqnarray}
Here,
$|\phi^\pm\rangle^P_{AC}=\frac{1}{\sqrt{2}}(|RR\rangle\pm|LL\rangle)_{AC}$,
$|\psi^\pm\rangle^P_{AC}=\frac{1}{\sqrt{2}}(|RL\rangle\pm|LR\rangle)_{AC}$,
and
$|\pm\rangle_{e_1}=\frac{1}{\sqrt{2}}(|-1\rangle\pm|+1\rangle)_{e_1}$.
The same polarization operations are performed on the two spatial
modes $i_1$ and $i_2$ ($i=a,c$) without affecting the states of the
photons in the spatial-mode DOF. By measuring the state of NV$_1$ in
the orthogonal basis $\{|+\rangle_{e_1}, |-\rangle_{e_1}\}$, we can
distinguish the even-parity polarization Bell states
($|\phi^\pm\rangle^P_{AC}$) from the odd-parity polarization Bell
states ($|\psi^\pm\rangle^P_{AC}$). That is,  the two-photon system
is in an even-parity polarization Bell state if NV$_1$ is projected
into the state $|+\rangle_{e_1}$. Otherwise, the two-photon system
is in an odd-parity polarization Bell state.

(2) \emph{Spatial-mode parity-check QND.} The spatial-mode
parity-check QND (S-QND) is constructed with NV$_2$ and HWPs shown
in Fig.\ref{figure2} (b). If we have a photon in the state
$\alpha|R\rangle+\beta|L\rangle$ put into NV$_2$ and HWP$_1$, after
interaction, the state of the system composed of NV$_2$ and the
photon is transformed into
\begin{eqnarray}                 
(\alpha|R\rangle+\beta|L\rangle) \otimes |+\rangle_{e_2}\rightarrow
(\alpha|R\rangle+\beta|L\rangle) \otimes |-\rangle_{e_2}.
\end{eqnarray}
By measuring the state of NV$_2$ in the orthogonal basis
$\{|+\rangle_{e_2}, |-\rangle_{e_2}\}$, we can distinguish the case
with an even number of photons ($|+\rangle_{e_2}$) from that with an
odd number of photons ($|-\rangle_{e_2}$) which  have interacted
with NV$_2$. Therefore, we can distinguish the even-parity
spatial-mode Bell states
$\left[|\phi^\pm\rangle^S_{AC}=\frac{1}{\sqrt{2}}(|a_1c_1\rangle\pm|a_2c_2\rangle)_{AC}\right]$
from the odd-parity spatial-mode Bell states
$\left[|\psi^\pm\rangle^S_{AC}=\frac{1}{\sqrt{2}}(|a_1c_2\rangle\pm|a_2c_1\rangle)_{AC}\right]$
by putting the two spatial modes $a_2$ and $c_2$ into NV$_2$ and HWP
(HWP$_1$ and HWP$_2$) in sequence, without influencing the states of
the photons in the polarization DOF.

(3) \emph{Measurement on an NV-center system.} By applying a
Hadamard operation on NV$_1$, the states $|+\rangle_{e_1}$ and
$|-\rangle_{e_1}$ can be rotated to the states $|-1\rangle_{e_1}$
and $|+1\rangle_{e_1}$ respectively. If we have an auxiliary photon
$p$
$\left[\vert\varphi_p\rangle=\frac{1}{\sqrt{2}}(|R\rangle+|L\rangle)\right]$
put into CPBS$_1$, NV$_1$, and CPBS$_3$ as shown in Fig.
\ref{figure2} (a), after interaction, the state of the system
composed of  NV$_1$ and the photon $p$ is changed as follows:
\begin{eqnarray}                            
\frac{1}{\sqrt{2}}(|R\rangle+|L\rangle)|-1\rangle_{e_1}   &\rightarrow&  \frac{1}{\sqrt{2}}(|R\rangle+|L\rangle)|-1\rangle_{e_1},\nonumber\\
\frac{1}{\sqrt{2}}(|R\rangle+|L\rangle)|+1\rangle_{e_1}
 &\rightarrow&
\frac{-1}{\sqrt{2}}(|R\rangle-|L\rangle)|+1\rangle_{e_1}.
\end{eqnarray}
We can read out the state of NV$_1$ by measuring the output state of
the auxiliary photon $p$ with orthogonal linear polarization basis.
If the auxiliary photon $p$ is in the state
$\frac{1}{\sqrt{2}}(|R\rangle+|L\rangle)$, the state of the NV
center is $|-1\rangle_{e_1}$. Otherwise, the state of the NV center
is $|+1\rangle_{e_1}$. The state of NV$_2$ can be measured in the
same way.

\subsection*{\uppercase\expandafter{2.2}  Hyper-EPP for a mixed hyperentangled Bell state with parity-check QNDs} \label{sec22}

We suppose that there are two identical two-photon systems in a
mixed hyperentangled state as follows:
\begin{eqnarray}                           
\rho_{AB}&=&\left[F_1|\phi^+\rangle_P\langle\phi^+|+(1-F_1)|\psi^+\rangle_P\langle\psi^+|\right]_{AB}\nonumber\\
&&\otimes\left[F_2|\phi^+\rangle_S\langle\phi^+|+(1-F_2)|\psi^+\rangle_S\langle\psi^+|\right]_{AB},\nonumber\\
\rho_{CD}&=&\left[F_1|\phi^+\rangle_P\langle\phi^+|+(1-F_1)|\psi^+\rangle_P\langle\psi^+|\right]_{CD}\nonumber\\
&&\otimes\left[F_2|\phi^+\rangle_S\langle\phi^+|+(1-F_2)|\psi^+\rangle_S\langle\psi^+|\right]_{CD}.\;\;\;\;
\end{eqnarray}
Here the subscripts $AB$ and $CD$ represent two photon pairs shared
by the two remote parties in quantum communication, say Alice and
Bob, respectively. The two photons $A$ and $C$ belong to Alice, and
the two photons $B$ and $D$ belong to Bob. $F_1$ and $F_2$ represent
the probabilities of $|\phi^+\rangle_P$ and $|\phi^+\rangle_S$,
respectively. The subscripts $P$ and $S$ represent the polarization
and the spatial-mode DOFs of  a two-photon system, respectively. The
initial state of the four-photon system $ABCD$ is
$\rho_0=\rho_{AB}\otimes\rho_{CD}$. It can be viewed as the mixture
of 16 maximally hyperentangled pure states.

The principle of our hyper-EPP for a mixed hyperentangled Bell state
with both polarization bit-flip errors and spatial-mode bit-flip
errors is shown in Fig.\ref{figure2} (c). Both Alice and Bob perform
the same parity-check QNDs on the polarization and the spatial-mode
DOFs of photon pairs $AC$ and $BD$, and the entanglement of a pair
of two-photon systems in the two DOFs can be purified independently.
For describing the principle of our hyper-EPP explicitly, we
describe the principle of our spatial-mode EPP in detail, and the
principle of our polarization EPP is the same as our spatial-mode
EPP.

The spatial-mode state
$|\phi^+\rangle^S_{AB}\otimes|\phi^+\rangle^S_{CD}$ is described as
\begin{eqnarray}                          
|\phi^+\rangle^S_{AB}\otimes|\phi^+\rangle^S_{CD}&=&\frac{1}{2}(|a_1b_1c_1d_1\rangle
+|a_2b_2c_2d_2\rangle\nonumber\\
&&+|a_1b_1c_2d_2\rangle+|a_2b_2c_1d_1\rangle).\;\;\;\;
\end{eqnarray}
Alice and Bob first perform the  S-QNDs on the two photon pairs $AC$
and $BD$, respectively. If the two photon pairs are both in either
an even-parity or odd-parity spatial mode, the spatial-mode state of
the four-photon system $ABCD$ is projected into
$|\Phi\rangle_1=\frac{1}{\sqrt{2}}(|a_1b_1c_1d_1\rangle+|a_2b_2c_2d_2\rangle)$
or
$|\Phi\rangle_2=\frac{1}{\sqrt{2}}(|a_1b_1c_2d_2\rangle+|a_2b_2c_1d_1\rangle)$,
respectively. The state $|\Phi\rangle_2$ can be transformed into
$|\Phi\rangle_1$ by performing spatial-mode bit-flip operations on
the photons $C$ and $D$. Then Alice and Bob perform Hardmard
operations on the spatial-mode DOF of the two photons $C$ and $D$,
respectively (with 50:50 BSs), and the state $|\Phi\rangle_1$ is
transformed into
\begin{eqnarray}                           
|\Phi'_1\rangle&=&\frac{1}{2\sqrt{2}}[(|a_1b_1\rangle+|a_2b_2\rangle)(|c_1d_1\rangle+|c_2d_2\rangle)\nonumber\\
&&+(|a_1b_1\rangle-|a_2b_2\rangle)(|c_1d_2\rangle+|c_2d_1\rangle)].\label{eq11}
\end{eqnarray}
If the two clicked photon detectors of photons $C$ and $D$ are in
the  even-parity spatial mode, the two-photon system $AB$ is
projected into the maximally entangled state
$|\phi^+\rangle^S_{AB}=\frac{1}{\sqrt{2}}(|a_1b_1\rangle+|a_2b_2\rangle)_{AB}$.
If the outcome of two clicked detectors is in the odd-parity spatial
mode, a phase-flip operation $\sigma^s_z=|b_1\rangle\langle b_1| -
|b_2\rangle\langle b_2|$ performed on the photon $B$ is required to
obtain the state $|\phi^+\rangle^S_{AB}$.

The spatial-mode states
$|\phi^+\rangle^S_{AB}\otimes|\psi^+\rangle^S_{CD}$ and
$|\psi^+\rangle^S_{AB}\otimes|\phi^+\rangle^S_{CD}$ are described as
\begin{eqnarray}                          
|\phi^+\rangle^S_{AB}\otimes|\psi^+\rangle^S_{CD}&=&\frac{1}{2}(|a_1b_1c_1d_2\rangle
+|a_2b_2c_2d_1\rangle\nonumber\\\
&&+|a_1b_1c_2d_1\rangle+|a_2b_2 c_1d_2\rangle),\nonumber\\
|\psi^+\rangle^S_{AB}\otimes|\phi^+\rangle^S_{CD}&=&\frac{1}{2}(|a_1b_2c_1d_1\rangle
+|a_2b_1c_2d_2\rangle\nonumber\\\
&&+|a_1b_2c_2d_2\rangle+|a_2b_1c_1d_1\rangle).\;\;\;\;\;\;\;\;
\end{eqnarray}
If the spatial-mode states of the two photon pairs $AC$ and $BD$ are
in the even-parity mode and the odd-parity mode, respectively, the
state of the four-photon system $ABCD$ is projected into
$|\Phi\rangle_3=\frac{1}{\sqrt{2}}(|a_1b_1c_1d_2\rangle+|a_2b_2c_2d_1\rangle)$
or
$|\Phi\rangle_4=\frac{1}{\sqrt{2}}(|a_1b_2c_1d_1\rangle+|a_2b_1c_2d_2\rangle)$.
As Alice and Bob cannot identify which one of the photon pairs $AB$
and $CD$ has bit-flip errors,   the two photon pairs have to be
discarded in this case. The two photon pairs also have to be
discarded if the spatial-mode states of the two photon pairs $AC$
and $BD$ are in the odd-parity mode and the even-parity mode,
respectively.

The spatial-mode state
$|\psi^+\rangle^S_{AB}\otimes|\psi^+\rangle^S_{CD}$ is described as
\begin{eqnarray}                          
|\psi^+\rangle^S_{AB}\otimes|\psi^+\rangle^S_{CD}&=&\frac{1}{2}(|a_1b_2c_1d_2\rangle
+|a_2b_1c_2d_1\rangle\nonumber\\
&&+|a_1b_2c_2d_1\rangle+|a_2b_1c_1d_2\rangle).\;\;\;\;\;\;\;\;
\end{eqnarray}
The spatial-mode states of the two photon pairs $AC$ and $BD$ are
both in either the even-parity mode or the odd-parity mode, which is
the same as $|\phi^+\rangle^S_{AB}\otimes|\phi^+\rangle^S_{CD}$.
Therefore, Alice and Bob cannot distinguish the case in which both
the two photon pairs with bit-flip errors from that without bit-flip
errors.

The principle of our polarization EPP is the same as our
spatial-mode one. That is, Alice and Bob pick up the cases in which
the two photon pairs $AC$ and $BD$ are in the same polarization
parity mode and discard the ones in different polarization parity
modes.

\begin{figure}[!h]
\centering
\includegraphics[width=8 cm,angle=0]{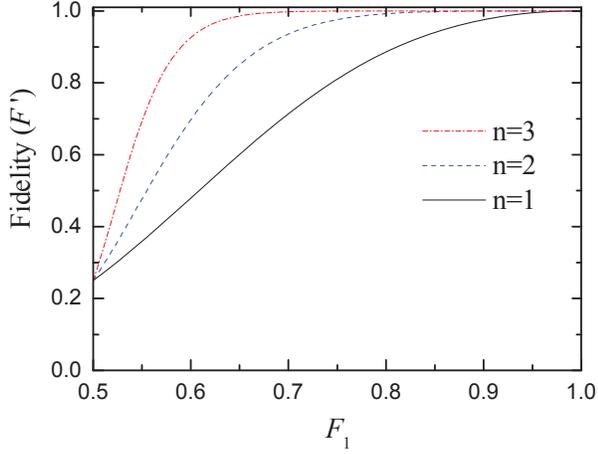}
\caption{ The fidelity of the two-photon hyperentangled Bell state
in our hyper-EPP. $F'$ alters with the iteration number of
hyperentanglement purification processes $n$ and the initial
fidelity of the mixed hyperentangled Bell state $F_1$. Here, we
assume $F_1=F_2$. } \label{figure3}
\end{figure}

After a round of our hyper-EPP process, the state of the photon pair
$AB$ kept is transformed into
\begin{eqnarray}                           
\rho'_{AB}&=&\left[F'_1|\phi^+\rangle_P\langle\phi^+|+(1-F'_1)|\psi^+\rangle_P\langle\psi^+|\right]_{AB} \nonumber\\
&&\otimes\left[F'_2|\phi^+\rangle_S\langle\phi^+|+(1-F'_2)|\psi^+\rangle_S\langle\psi^+|\right]_{AB}.\;\;\;\;\;\;\;\;\;\label{eq23}
\end{eqnarray}
Here $F'_1=\frac{F_1^2}{[F_1^2+(1-F_1)^2]}$,
$F'_2=\frac{F_2^2}{[F_2^2+(1-F_2)^2]}$, and $F_i>1/2$ ($i=1,2$). The
fidelity of $|\phi^+\rangle^P_{AB}|\phi^+\rangle^S_{AB}$ in
Eq.(\ref{eq23}) is $F'=F'_1\times F'_2$. With the iteration of our
hyper-EPP process several times, the fidelity of this two-photon
hyperentangled Bell state can be improved (shown in Fig.
\ref{figure3} for the cases with $F_1=F_2$).

Our hyper-EPP is constructed to purify a mixed hyperentangled Bell
state with bit-flip errors in both the spatial-mode and the
polarization DOFs. Usually, the two spatial modes of a photon are
composed of two fibers (pathes).  In fact, the spatial-mode state of
photon pairs is far more robust than their polarization state in a
practical transmission over optical fiber channels
\cite{EPPsimon,EPPexperiment,EPPsheng3,EPPLi,EPPdeng1}.  Especially,
the probability  that  the  bit-flip  error  takes  place  in  the
spatial-mode DOF of a photon pair is negligible as it is difficult
for a photon to penetrate from a fiber to the other. However, the
phases of two wavepackets for a photon are not stable as they are
influenced by the thermal fluctuation, vibration, and the
imperfection of the fibers \cite{APL}. In principle, the phase-flip
errors can be transformed into bit-flip errors with local Hadamard
operations in both the spatial-mode and the polarization DOFs
\cite{EPPsheng1,EPP3,EPPsimon,EPPexperiment,HECP}. Therefore, our
hyper-EPP can also be used to purify the mixed hyperentangled Bell
states with phase-flip errors in two DOFs. If the two spatial modes
of a photon $i$ in a mixed hyperentangled Bell state have different
polarization errors, such as a polarization bit-flip error in $i_1$
and a polarization phase-flip error in $i_2$, the polarization
phase-flip error can be transformed into a polarization bit-flip
error with local operation, and this mixed hyperentangled Bell state
can also be purified with our hyper-EPP.

\section*{\uppercase\expandafter{3.} Hyper-ECP for a partially
hyperentangled  Bell  state with  cavity-NV-center systems}

\label{sec3}

\begin{figure}[!h]
\centering\includegraphics[width=7.0 cm,angle=0]{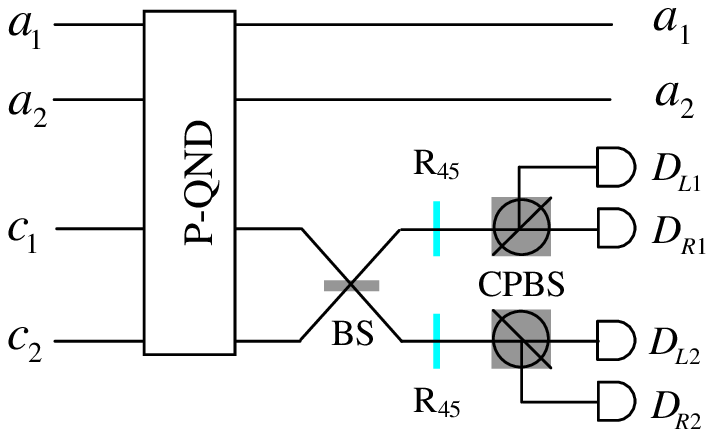} \caption{
Schematic diagram of our hyper-ECP for a partially hyperentangled
Bell  state with unknown parameters, resorting to nonlinear optics.
Bob performs the same operations as Alice by replacing the photons
$AC$ and P-QND with the photons $BD$ and S-QND, respectively.  }
\label{figure4}
\end{figure}

We suppose that there are two identical two-photon systems in a
nonlocal partially hyperentangled Bell  state in both the
polarization and spatial-mode DOFs, that is,
\begin{eqnarray}                           
|\psi_0\rangle_{AB}&=&(\alpha|RR\rangle+\beta|LL\rangle)_{AB}
\otimes(\gamma|a_1b_1\rangle+\delta|a_2b_2\rangle),\nonumber\\
|\psi_0\rangle_{CD}&=&(\alpha|RR\rangle+\beta|LL\rangle)_{CD}
\otimes(\gamma|c_1d_1\rangle+\delta|c_2d_2\rangle).\nonumber\\
\end{eqnarray}
Here, the subscripts $AB$ and $CD$ represent two photon pairs shared
by Alice and Bob, respectively. The two photons $A$ and $C$ are kept
by Alice, and the two photons $B$ and $D$ are kept by Bob. $\alpha$,
$\beta$, $\gamma$, and $\delta$ are four real parameters that are
unknown to Alice and Bob, and they satisfy the relation
$|\alpha|^2+|\beta|^2=|\gamma|^2+|\delta|^2=1$.

The principle of our hyper-ECP for a nonlocal partially
hyperentangled Bell-class state with unknown parameters is shown in
Fig. \ref{figure4}. The initial state of the four-photon system
$ABCD$ is
$|\Psi_0\rangle=|\psi_0\rangle_{AB}\otimes|\psi_0\rangle_{CD}$.
Alice can divide the states of the two-photon system $AC$ into two
groups with the P-QND whose principle is shown in Fig. \ref{figure2}
(a), and Bob can divide the states of the two-photon system $BD$
into two groups with the S-QND shown in Fig. \ref{figure2} (b). They
pick up the odd-parity terms of the polarization DOF and the
odd-parity terms of the spatial-mode DOF with the same parameters,
respectively. The state of the four-photon system with the selected
terms is
\begin{eqnarray}                           
|\Psi_1\rangle&=&\frac{1}{2}(|RRLL\rangle+|LLRR\rangle)_{ABCD}\nonumber\\
&& \otimes (|a_2b_2c_1d_1\rangle+|a_1b_1c_2d_2\rangle).\label{eq10}
\end{eqnarray}
The probability for obtaining this state is
$p(1)=4|\alpha\beta\gamma\delta|^2$.  Alice and Bob can obtain the
photon pair $AB$ in the maximally hyperentangled Bell state
$|\phi^+\rangle^P_{AB}|\phi^+\rangle^S_{AB}$ with Hadamard
operations, detections, and local phase-flip operations on this
four-photon system, as the same as that discussed in our hyper-EPP.

If the state of the four-photon system $ABCD$ is in an even-parity
polarization mode and an even-parity spatial mode, the two-photon
system $AB$ is projected into $|\psi_1\rangle_{AB}$ after the
detections on two photons $CD$ and local phase-flip operations on
$B$. Here
\begin{eqnarray}                           
|\psi_1\rangle_{AB}&=&y(\alpha^2|RR\rangle+\beta^2|LL\rangle)_{AB}\nonumber\\
&&\otimes
(\gamma^2|a_1b_1\rangle+\delta^2|a_2b_2\rangle),\label{eq12}
\end{eqnarray}
and
$y=\frac{1}{\sqrt{(|\alpha|^4+|\beta|^4)(|\gamma|^4+|\delta|^4)}}$.
This is a hyperentangled state with less entanglement, which takes
place with the probability of
$p'(1)_1=(|\alpha|^4+|\beta|^4)(|\gamma|^4+|\delta|^4)$, and it can
be distilled to the maximally hyperentangled Bell state with another
round of our hyper-ECP. That is, another photon pair $A'B'$ is
required, which has the identical nonlocal partially hyperentangled
Bell state with the photon pair $AB$. Alice and Bob perform P-QND
and S-QND on the photon pairs $AA'$ and $BB'$, respectively, and
they pick up the odd-parity terms of the polarization DOF and the
odd-parity terms of the spatial-mode DOF with the same parameters,
respectively. In this way,  the maximally hyperentangled Bell state
can be obtained with the probability of
$p(2)_1=4|\alpha\beta\gamma\delta|^4/[(|\alpha|^4+|\beta|^4)(|\gamma|^4+|\delta|^4)]$
after Hadamard operations, detections, and local phase-flip
operations. If the state of the four-photon system $ABA'B'$ is not
in an odd-parity polarization mode and an odd-parity spatial mode, a
third round of hyper-ECP process is required.

In the other two cases that the outcomes of the two DOFs are in
different parity modes, there is one DOF of the two-photon system
$AB$ projected into the maximally entangled Bell state. That is,
\begin{eqnarray}                           
|\psi_3\rangle_{AB}&=&y_1(|RR\rangle+|LL\rangle)_{AB}
\otimes(\gamma^2|a_1b_1\rangle+\delta^2|a_2b_2\rangle),\nonumber\\
|\psi_4\rangle_{AB}&=&y_2(\alpha^2|RR\rangle+\beta^2|LL\rangle)_{AB}
\otimes(|a_1b_1\rangle+|a_2b_2\rangle),\nonumber\\
\end{eqnarray}
where $y_1=\frac{1}{\sqrt{2(|\gamma|^4+|\delta|^4)}}$ and
$y_2=\frac{1}{\sqrt{2(|\alpha|^4+|\beta|^4)}}$. These two states are
obtained with the probabilities of
$p'(1)_2=2|\alpha\beta|^2(|\gamma|^4+|\delta|^4)$ and
$p'(1)_3=2|\gamma\delta|^2(|\alpha|^4+|\beta|^4)$, respectively. In
a second round of hyper-ECP process, the maximally hyperentangled
Bell state can be obtained with the probabilities of
$p(2)_2=4|\gamma\delta|^4|\alpha\beta|^2/(|\gamma|^4+|\delta|^4)$
and
$p(2)_3=4|\alpha\beta|^4|\gamma\delta|^2/(|\alpha|^4+|\beta|^4)$,
respectively. If the state of the four-photon system $ABA'B'$ is not
in an odd-parity spatial mode (similar to $|\psi_3\rangle_{AB}$) or
an odd-parity polarization mode (similar to $|\psi_4\rangle_{AB}$),
a third round of hyper-ECP process is required.

The success probability $P$ of our hyper-ECP becomes much higher by
the iteration of our hyper-ECP process with the states preserved in
the latter three cases, which have been discarded in the hyper-ECP
with linear optics \cite{HECP}. That is, the success probabilities
of each round are $p(1)$, $p(2)=p(2)_1+p(2)_2+p(2)_3$, \ldots,
respectively. The total success probability of our hyper-ECP with
$n$ rounds is $P=\sum_np(n)$.

\begin{figure}[htb]                    
\centering
\includegraphics[width=8 cm]{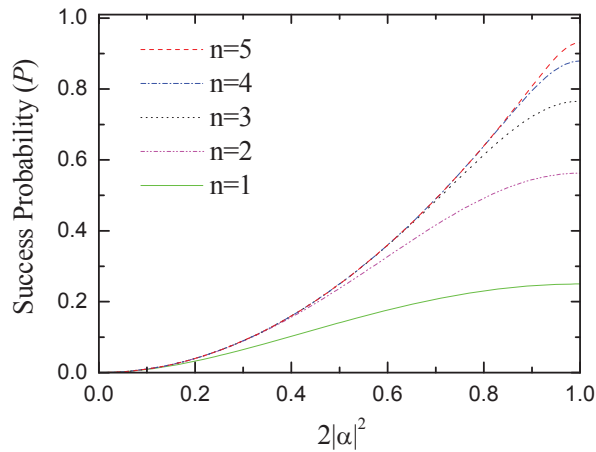}
\caption{ Success probability $P$ of our hyper-ECP for a partially
hyperentangled Bell state under the iteration numbers of hyper-ECP
process $n$. The parameters of the partially hyperentangled Bell
state are chosen as $|\alpha|=|\gamma|$ and $|\beta|=|\delta|$.}
\label{figure5}
\end{figure}

In the case $|\alpha|<|\beta|$ ($|\alpha|=|\gamma|$ and
$|\beta|=|\delta|$), the relation between the success probability
$P$ and the parameter 2$|\alpha|^2$ is shown in Fig. \ref{figure5}.
If the parameter is 2$|\alpha|^2=0.9$, the success probability of
our hyper-ECP is $P=80.8\%$ with the iteration number $n=5$ (nearly
equivalent to the maximal value $4|\alpha|^4$=81\%), while it is
$P=24.5\%$ for the hyper-ECP with linear optics \cite{HECP}. That
is, the success probability $P$ of our hyper-ECP can achieve the
maximal success probability in principle.

\section*{\uppercase\expandafter{4.}  Discussion and summary}
\label{sec4}

An NV center in a diamond is a promising candidate for quantum
information processing with its long electron-spin decoherence time
even at the room temperature \cite{NV,NV1}, and it has nanosecond
manipulation time for an individual NV center \cite{NV6}. The
interaction between an NV center and a circularly polarized photon
may be largely enhanced by coupling with the photonic crystal cavity
\cite{CNV5}. Therefore, the fidelities of the P-QND and the S-QND in
our hyper-EPP and hyper-ECP are mainly influenced by the coupling
strength and cavity side leakage. With the definition
$F=|\langle\psi_f|\psi\rangle|^2$, we can obtain the fidelities of
the two QNDs  (for even parity states),
\begin{eqnarray}                           \label{eq.17}
F_P&=&\frac{(|r_0|^2+|r|^2+2)^2(|r_0|+|r|+2)^2}{16(|r_0|^4+|r|^4+2)(|r_0|^2+|r|^2+2)},\nonumber\\
F_S&=&\left[\frac{1}{2}+\frac{(|r_0|^2+|r|^2)^2}{4(|r_0|^4+|r|^4)}\right]\frac{(|r_0|+|r|+2)^2}{4(|r_0|^2+|r|^2+2)}.\;\;\;\;\;\;\;\;
\end{eqnarray}
Here $|\psi\rangle$ is the ideal finial state and $|\psi_f\rangle$
is the finial state considering experimental factors. The
spontaneous decay rate of an NV center is $\gamma=2\pi\times15$ MHZ
\cite{NV2}, and the cavity side leakage rate can reach
$10\kappa=\eta=2\pi\times10$GHZ \cite{CNV} with a quality factor of
$Q\approx10^4$. If the coupling strength is $g=0.1\eta$, the
fidelities of these two QNDs are $F_P=97.1\%$ and $F_S=98.6\%$,
respectively. That is, both the P-QND and the S-QND are feasible
with current technology.

With linear optics, we can only purify one qubit error for a pair of
two-photon systems in a nonmaximally hyperentangled state. That is,
a pair of two-photon systems in a mixed hyperentangled Bell state
with both polarization bit-flip errors and spatial-mode bit-flip
errors cannot be purified with linear optics \cite{HECP}. Therefore,
nonlinear optics is required to implement the hyper-EPP for a mixed
hyperentangled Bell state with two or more qubit errors in two DOFs.
Our hyper-EPP is the first one for a mixed hyperentangled
spatial-polarization Bell state with the errors in both the two DOFs
with the P-QNDs and S-QNDs assisted by one-side cavity-NV-center
systems. As the phase-flip errors in both the polarization and the
spatial-mode DOFs can be transformed into the bit-flip errors in
these two DOFs \cite{EPPsheng1,EPP3}, respectively, our hyper-EPP
can be used to purify a mixed hyperentangled Bell state completely
in principle. Besides, we have proposed an efficient hyper-ECP for
partially hyperentangled Bell  states with unknown parameters,
resorting to nonlinear optics. Our hyper-ECP can obtain the maximal
success probability in principle by the iteration of the hyper-ECP
process with the states that have been discarded in the hyper-ECP
with linear optics \cite{HECP}.

In summary, we have investigated the possibility of improving the
entanglement of nonlocal nonmaximally hyperentangled Bell states
with nonlinear optics. Both our hyper-EPP and hyper-ECP are
implemented with the P-QNDs and S-QNDs that are constructed with
nonlinear optics of cavity-NV-center systems. We have analyzed the
experimental feasibility of the two QNDs in two DOFs and shown that
they can be implemented with current technology. These two protocols
are useful for increasing the channel capacity of long-distance
quantum communication with hyperentanglement.

\section*{Acknowledgments}

This work is supported by the National Natural Science Foundation of
China under Grant No. 11174039 and NECT-11-0031.

\end{document}